\documentclass[iop,usenatbib]{emulateapj}

\usepackage[english]{babel}
\usepackage{graphicx}
\usepackage{fleqn}
\usepackage{amssymb}
\usepackage{amsmath}
\usepackage{booktabs}
\usepackage{hyperref}

\renewcommand{\S}{Section}
\newcommand{\F}{Fig.}

\begin{document}

\title{Hints for hidden planetary companions to hot Jupiters in stellar binaries}

\author{Adrian S. Hamers}
\affil{Institute for Advanced Study, School of Natural Sciences, Einstein Drive, Princeton, NJ 08540, USA}
\email{hamers@ias.edu}

\begin{abstract} 
Searches for stellar companions to hot Jupiters (HJs) have revealed that planetary systems hosting a HJ are approximately three times more likely to have a stellar companion with a semimajor axis between 50 and 2000 AU, compared to field stars. This correlation suggests that HJ formation is affected by the stellar binary companion. A potential model is high-eccentricity migration, in which the binary companion induces high-eccentricity Lidov-Kozai (LK) oscillations in the proto-HJ orbit, triggering orbital migration driven by tides. A pitfall of this `binary-LK' model is that the observed stellar binaries hosting HJs are typically too wide to produce HJs in sufficient numbers, because of suppression by short-range forces. We propose a modification to the binary-LK model in which there is a second giant planet orbiting the proto-HJ at a semimajor axis of several tens of AU. Such companions are currently hidden to observations, but their presence could be manifested by a propagation of the perturbation of the stellar binary companion inwards to the proto-HJ, thereby overcoming the barrier imposed by short-range forces. Our model does not require the planetary companion orbit to be eccentric and/or inclined with respect to the proto-HJ, but its semimajor axis should lie in a specific range given the planetary mass and binary semimajor axis, and the inclination with respect to the binary should be near $40^\circ$ or $140^\circ$. Our prediction for planetary companions to HJs in stellar binaries should be testable by future observations.
\end{abstract}

\section{Introduction}
\label{sect:introduction}
Recent adaptive optics (AO) searches of stellar companions to hot Jupiters (HJs) \citep{2014ApJ...785..126K,2015ApJ...800..138N,2015ApJ...814..148P,2016ApJ...827....8N} have shown that the presence of a HJ in the planetary system is related to the binary fraction. In particular, in the sample of \citet{2016ApJ...827....8N}, planetary systems hosting HJs are $\approx 3$ times more likely to have a stellar companion with a semimajor axis between 50 and 2000 AU compared to field stars. In contrast, systems with HJs are $\sim 2$ to 8 times less likely to have a stellar companion in the range 1-50 AU compared to field stars. This suggests that the presence of a stellar companion is closely related to giant planet formation and/or planetary migration processes.

One considered planetary migration process is high-eccentricity or high-$e$ migration. In this model, the binary companion induces secular Lidov-Kozai (LK) oscillations \citep{1962P&SS....9..719L,1962AJ.....67..591K}, periodically exciting high eccentricities of the giant planet orbit, which is initially thought to have a semimajor axis of a few AU \citep{2003ApJ...589..605W,2007ApJ...669.1298F,2012ApJ...754L..36N,2015ApJ...799...27P,2016MNRAS.456.3671A,2016ApJ...829..132P}. The associated small pericenter distances trigger strong tidal dissipation, and the planetary orbit is circularized with an orbital period of a few days. 

The `binary-LK' model is challenged by recent observations of population statistics of stellar companions. The orbits of the stellar companions are typically too wide to induce LK oscillations in the planetary orbit because of quenching by short-range forces, notably general relativistic precession \citep{2016ApJ...827....8N}. This suggests that binary-LK migration was not active in the majority of observed HJ systems with stellar companions, and the observed correlation of HJ occurrence with stellar binary companions may be due to a another phenomenon, e.g. a relation between giant planet and stellar binary companion formation \citep{2016ApJ...827....8N}. 

Current observations, however, do not exclude the possibility of giant planet or substellar companions to HJs in the semimajor axis range $\sim$ 10-50 AU \citep{2016ApJ...821...89B}. Such more distant companions could have formed around the disk of the primary star e.g. through gravitational instability, given that the typical truncation radius of the protoplanetary disk by the stellar binary companion is $\sim 1/3$ of the binary semimajor axis \citep{1999AJ....117..621H}, and the latter is typically several hundred AU in the sample of \citet{2016ApJ...827....8N}. If present, such a companion could affect the dynamics of the binary-LK model.

In this paper, we investigate the dynamical effect of a second planet in the binary-LK model, assuming that the proto-HJ initially formed at a few AU. We will show that HJs can be produced through high-$e$ migration in stellar binaries that are too wide to drive HJ migration in the absence of a second planet. The underlying mechanism is that the stellar binary companion can make the orbit of the second planet eccentric and inclined with respect to the orbit of the proto-HJ. Subsequently, the second planet can excite high-eccentricity LK-oscillations in the orbit of the proto-HJ, triggering high-$e$ migration. 

Our model does not require the orbit of the second planet to be initially inclined with respect to the proto-HJ, nor does it have to be initially eccentric. Coplanarity would be expected if these objects formed out of the same disk (assuming the latter was not warped by the binary companion, e.g. \citealt{2014ApJ...792L..33M}). A requirement for our model is that the orbit of the second planet has a specific range of semimajor axes, given its mass and the stellar binary semimajor axis. Also, the inclination of the second planet with respect to the stellar binary should be near $40^\circ$ or $140^\circ$. We will show that a second planet can strongly enhance the HJ formation rate, although the above requirements imply that the parameter space for HJ enhancement is small.

\section{Methodology}
\label{sect:meth}

\begin{figure}
\centering
\includegraphics[scale = 0.5, trim = 0mm 8mm 0mm 0mm]{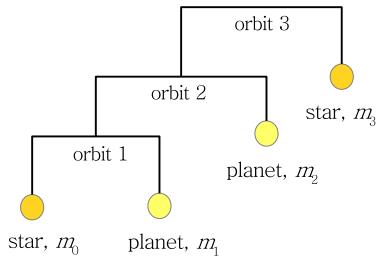}
\caption{\small Schematic representation of the two-planet system in a stellar binary considered in this paper.  }
\label{fig:configurations.eps}
\end{figure}

The systems studied here are S-type two-planet systems in stellar binaries, which can be considered as `3+1' hierarchical quadruple systems. The structure and notations are illustrated in \F\,\ref{fig:configurations.eps}. The long-term secular evolution is modeled using the code of \citet{2015MNRAS.449.4221H}, with updates from \citet{2016MNRAS.459.2827H}. The method assumes that the system is sufficiently hierarchical, and evolves the orbital orientations and eccentricities, assuming that the semimajor axes are constant (except for tidal evolution). The hierarchy condition can be violated, in particular if the orbit of the second planet is excited to high eccentricity and becomes unstable with respect to the proto-HJ orbit. To take such events into account, the simulations were stopped if orbit 1 became unstable with respect to orbit 2 using the dynamical stability criterion of \citet{2015ApJ...808..120P} (suitable for two-planet systems), or if orbit 2 became unstable with respect to orbit 3 using the dynamical stability criterion of \citet{1999AJ....117..621H} (suitable for test masses in S-type orbits in binaries). Relativistic precession was included to the first post-Newtonian order, neglecting interactions between binaries. The latter interactions may lead to further eccentricity excitation \citep{2013ApJ...773..187N}, but are beyond the scope of this paper.

We included tidal evolution in the proto-HJ and the primary star assuming the equilibrium tide model of \citet{1998ApJ...499..853E}, also taking into account their spin evolution (magnitude and direction -- the spin periods were set to $P_\mathrm{spin,0} = 10 \, \mathrm{d}$ and $P_\mathrm{spin,1} = 10 \, \mathrm{hr}$, and the initial obliquities were set to zero). Precession of the orbits due to tidal bulges and rotation was also taken into account. We assumed a constant viscous time-scale of $t_\mathrm{V,0}=100\,\mathrm{yr}$ for the primary star, corresponding to a tidal quality factor of $Q_0 \sim 6 \times 10^5$ for a HJ at 0.05 AU, typical for main-sequence stars (\citealt{2012MNRAS.423..486L}, and references therein). Regarding the proto-HJ,  \citet{2012arXiv1209.5724S} provided the constraint $t_\mathrm{V,1} < 1.4\, \mathrm{yr}$, by requiring that a HJ at 5 d is circularized in less than 10 Gyr; we assumed three values of the viscous time-scale, $t_\mathrm{V,1} \in \{0.014,0.14,1.4\} \, \mathrm{yr}$. The apsidal motion constants were set to $k_\mathrm{AM,0} = 0.014$ and $k_\mathrm{AM,1} = 0.25$, and the gyration radii ($I_i = r_\mathrm{g} m_i R_i^2$ where $I_i$ is moment of inertia and $R_i$ is the radius) to $r_\mathrm{g,0}=0.08$ and $r_\mathrm{g,1}=0.25$.

\section{Example system}
\label{sect:example}

\begin{figure}
\center
\includegraphics[scale = 0.43, trim = 10mm 5mm 10mm 20mm]{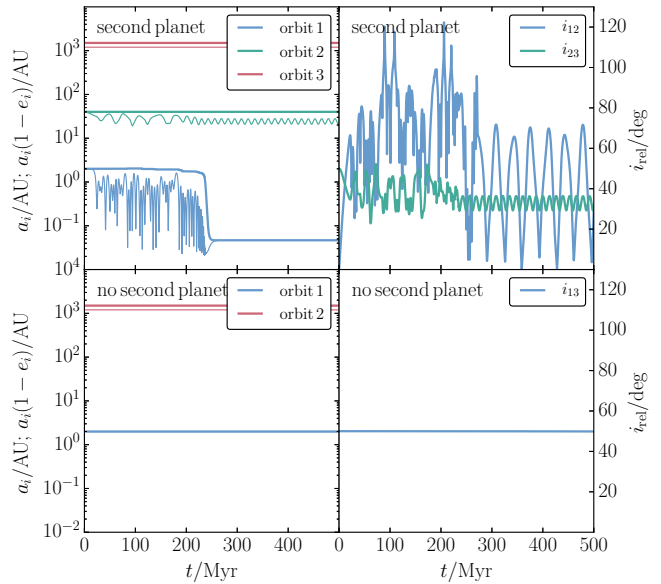}
\caption{\small Example evolution of a system in which high-$e$ migration is orchestrated by a second planet in a stellar binary system. Left column: semimajor axes (thick lines) and pericenter distances (thin lines) as a function of time. Right column: mutual inclinations as a function of time. Refer to the text for the parameters. The top (bottom) row shows the case with (without) the second planet. }
\label{fig:example1_test01_detailed.eps}
\end{figure}

An example is given in \F\,\ref{fig:example1_test01_detailed.eps}, with $m_0=1\,\mathrm{M}_\odot$, $m_1 = m_2 = 1\,M_\mathrm{J}$ and $m_3 = 0.4\,\mathrm{M}_\odot$. The parameters for the case without the second planet (second row) were chosen to represent a stellar binary ($a_3 = 1500\,\mathrm{AU}$ and $e_2=0.2$) that is not conducive to drive high-$e$ migration in a Jupiter-mass planet, originally in a circular orbit at $a_1 = 2\,\mathrm{AU}$. The LK time-scale without the second planet is $\sim 2800\,\mathrm{Myr}$, which is much longer than the relativistic precession time-scale, $t_\mathrm{GR,1} \sim 190\,\mathrm{Myr}$. In the figure shown, the initial mutual inclination $i_{13} = 50^\circ$ is modest; however, the same conclusion applies if $i_{13}$ were initially close to $90^\circ$. 

In the first row of \F\,\ref{fig:example1_test01_detailed.eps}, we included a second planet with $m_2=1\,M_\mathrm{J}$, $a_2 = 40\,\mathrm{AU}$ and $e_2 = 0.01$, initially coplanar with respect to the innermost planet ($i_{12}=0^\circ$). The evolution is now clearly different. The LK time-scales associated with the (1,2) and (2,3) orbit pairs are $\sim 24$ and 31 Myr, respectively, both shorter than the relativistic precession time-scale. Due to the secular torque of the stellar binary companion acting on the second planet, the orbit of the latter becomes inclined with respect to the innermost planet, inducing high $i_{12}$. Consequently, orbit 1 is excited to high eccentricity and small pericenter distances, triggering strong tidal dissipation. A HJ is formed after $\sim 250\,\mathrm{Myr}$ (for this example, we set $t_\mathrm{V,1} = 0.014\,\mathrm{yr}$). 

During the first $\sim 200 \,\mathrm{Myr}$ of the evolution, there is an indication for secular chaos in the system. In this process, high eccentricities can be attained in the orbits of inner planets due to overlap of secular time-scales \citep{2011ApJ...735..109W,2011ApJ...739...31L,2014PNAS..11112610L}. In this case, there is an overlap between the secular time-scales associated with the two planets, and with the secular driving of the second planet by the stellar binary companion. This is borne out by inspection of the angular momentum deficit, which tends to be equipartitioned among the two planets. 

The eccentricity of orbit 2 is slightly excited by the stellar binary companion. During the proto-HJ phase, the oscillations are periodic but somewhat irregular. After HJ formation, the oscillations become regular and with slightly smaller amplitude, showing that the second planet will remain in a stable orbit. 

In \F\,\ref{fig:example1_test01_detailed.eps}, initially $i_{13} = 50^\circ$. If $i_{13}$ were closer to $90^\circ$, orbit 2 would be excited to very high eccentricity, and the system would no longer be dynamically stable. This gives constraints on the orbital properties of the second planet to trigger HJ formation.

\section{Population synthesis}
\label{sect:pop_syn}

\subsection{Setup}
\label{sect:pop_syn:setup}
Our focus is on stellar binaries similar to the sample of \citet{2016ApJ...827....8N}, which are not conducive to HJ formation in the absence of a companion planet to the proto-HJ. The goal is to pinpoint the regions in parameter space where a second planet could efficiently drive high-$e$ migration of the innermost planet. We emphasize that our assumptions for the (currently unknown) second planet orbital parameter distributions, and thereby the HJ fractions, may not be representative of real systems. 

We sampled both $a_1$ and $a_2$ from flat distributions with the ranges $1\,\mathrm{AU}<a_1<5\,\mathrm{AU}$ and $10\,\mathrm{AU}<a_2<50\,\mathrm{AU}$. The lower limit on $a_2$ is justified a posteriori since no HJs are formed in the simulations for $a_2 \lesssim 20 \, \mathrm{AU}$. The upper limit on $a_2$ is motivated by the lower limit of $a_3 = 50 \, \mathrm{AU}$ from the sample of \citet{2016ApJ...827....8N}. The stellar binary orbital period was sampled from a lognormal distribution \citep{2010ApJS..190....1R}, with an upper cutoff at $10^{7.5}$ days (corresponding to $\sim 2000$ AU), matching the sample of \citet{2016ApJ...827....8N}. The initial eccentricities were set to $e_1=e_2 = 0.01$; $e_3$ was sampled from a Rayleigh distribution with an rms width of 0.33 between 0.01 and 0.9, approximating the distribution of \citet{2010ApJS..190....1R}. The masses were $m_0 = 1\,\mathrm{M}_\odot$, $m_1 = 1\,M_\mathrm{J}$; $m_2$ was sampled from a flat distribution between 1 and 10 $M_\mathrm{J}$ and $m_3$ was sampled from a flat distribution between 0.08 and 0.6 $\mathrm{M}_\odot$, the latter crudely approximating the mass ratio distribution of \citet{2016ApJ...827....8N} (assuming a primary mass of 1 $\mathrm{M}_\odot$). Sampled systems were rejected if they did not satisfy the requirements for dynamical stability (cf. \S\,\ref{sect:meth}). 

We assumed zero initial mutual inclination between orbits 1 and 2 ($i_{12}=0^\circ$), and a random orientation of orbits 1 and 2 with respect to orbit 3 (i.e. a flat distribution in $\cos i_{23}$). The arguments of pericenter $\omega_i$ and longitudes of the ascending nodes $\Omega_i$ were sampled from flat distributions. 

Each system was integrated for 10 Gyr, unless the innermost planet was tidally disrupted (assuming a disruption distance of $r_\mathrm{t} = \eta R_1(m_0/m_1)^{1/3}$ with $\eta = 2.7$, \citealt{2011ApJ...732...74G}) or if a dynamical instability occurred, in which case the secular integration was terminated. HJ systems were defined as systems with the innermost orbital period less than 10 d and an eccentricity less than $10^{-3}$. 

\begin{figure}
\center
\includegraphics[scale = 0.47, trim = 10mm 10mm 10mm 20mm]{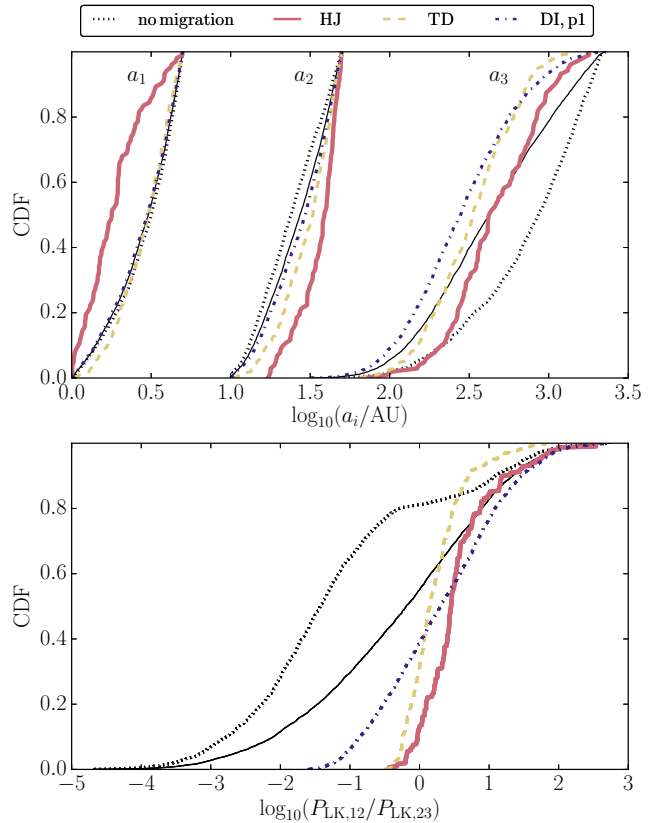}
\caption{\small Top panel: the initial distributions of the semimajor axes of the three orbits (annotated with $a_i$) arranged by the outcome of the simulations ($t_\mathrm{V,1} = 0.014\,\mathrm{yr}$). Distinguished are no migration (black dotted lines), HJ formation (red solid lines), tidal disruption (`TD'; yellow dashed lines) and dynamical instability of the orbit of the innermost planet (`DI, p1'; blue dot-dashed lines). The initial distributions for all sampled systems are shown with the thin black solid lines. Bottom panel: the initial distributions of the ratio of LK time-scales for orbit pairs (1,2) and (2,3) (cf. equation~\ref{eq:R}). }
\label{fig:arranged_sma_distributions_combined_test03.eps}
\end{figure}

\subsection{Results}
\label{sect:pop_syn:results}
With the inclusion of the second planet, the HJ fractions after 10 Gyr were approximately 0.024, 0.005 and 0.002 for inner planet viscous time-scales of $t_\mathrm{V,1} = 0.014$, 0.14 and 1.4 yr, respectively (fractions based on 1000 Monte Carlo realizations each). Without the second planet, these respective fractions were 0.052, 0.030 and 0.011. On face value, this would suggest that the second planet {\it decreases} the chances of producing a HJ. However, we emphasize that with the companion planet, a very large fraction ($\approx 0.5$) of systems became dynamically unstable, and the HJ fractions are strongly dependent on the assumed orbital parameters of the second planet. If the parameter range for the second planet were restricted, the HJ fractions would be strongly enhanced. 

To illustrate this, we show in the top panel of \F\,\ref{fig:arranged_sma_distributions_combined_test03.eps} the initial distributions of the semimajor axes arranged by the outcomes of the simulations with $t_\mathrm{V,1} = 0.014\,\mathrm{yr}$: no migration (black dotted lines), HJ formation (red solid lines), tidal disruption (`TD'; yellow dashed lines) and dynamical instability of the orbit of the innermost planet (`DI, p1'; blue dot-dashed lines). HJ systems show distinct distributions; $a_1$ is preferentially small, whereas $a_2$ should be at least $\sim 30\,\mathrm{AU}$. The minimum $a_3$ is $\sim 100\,\mathrm{AU}$; for smaller values, dynamical instabilities or tidal disruptions are more likely. 

These preferences for the semimajor axes are reflected in the ratio of LK time-scales of the orbital pairs (1,2) and (2,3), i.e. \citep{2015MNRAS.449.4221H}
\begin{align}
\label{eq:R}
\nonumber &\mathcal{R} \equiv P_\mathrm{LK,12}/P_\mathrm{LK,23} \\
&\sim \left ( \frac{a_2^3}{a_1 a_3^2} \right )^{3/2} \left ( \frac{m_0+m_1}{m_0+m_1+m_2} \right )^{1/2} \frac{m_3}{m_2} \left ( \frac{1-e_2^2}{1-e_3^2} \right )^{3/2}.
\end{align}
The initial distributions of this ratio, arranged by the outcomes, are shown in the bottom panel of \F\,\ref{fig:arranged_sma_distributions_combined_test03.eps}. HJ and tidal disruption systems show a strong preference for $\mathcal{R}$ close to unity. 

This result is expected based on the general property of quadruple systems that high eccentricities can be attained in the innermost orbit if $\mathcal{R}\sim 1$ \citep{2015MNRAS.449.4221H}. In the limit $\mathcal{R}\ll 1$, coplanarity between orbits 1 and 2 is maintained, even while their absolute inclinations change due to the torque of orbit 3. Due to precession induced on orbit 2 by orbit 1, orbit 2 is not excited in eccentricity by the torque of orbit 3. In the limit $\mathcal{R}\gg 1$, orbits 1 and 2 are decoupled, i.e. they do not remain coplanar as orbit 3 changes the orientation of orbit 2. There is no induced precession on orbit 2 by orbit 1, implying potentially high eccentricities in orbit 2, and potentially, dynamical instability. Note that physically, for small $i_{23}$ the precession time-scale $t_\mathrm{prec,23}$ of the angular momentum vector of orbit 2 around orbit 3 is the relevant time-scale to consider in the denominator of equation~(\ref{eq:R}), since in this case precession gives rise to a mutual inclination between orbits 1 and 2 rather than the LK mechanism. However, for the purpose of estimating the parameter space in which the innermost planet is driven to high eccentricity, it suffices to use the LK time-scale $P_\mathrm{LK,23}$ instead of a more accurate expression for $t_\mathrm{prec,23}$.

There is no strong dependence of the outcomes on the initial eccentricity $e_3$, nor on the masses $m_2$ and $m_3$. This can be understood qualitatively from equation~(\ref{eq:R}) by noting that the strongest dependence of $\mathcal{R}$ is on the semimajor axes. The semimajor axis distribution of the HJs (not shown here) is very similar to other alternative studies of high-$e$ migration (e.g. multiplanet systems, \citealt{2017MNRAS.464..688H}, or stellar triples, \citealt{2017arXiv170101733H}), and consistent with the observations of \citet{2016A&A...587A..64S}.

\begin{figure}
\center
\includegraphics[scale = 0.45, trim = 0mm 10mm 10mm 20mm]{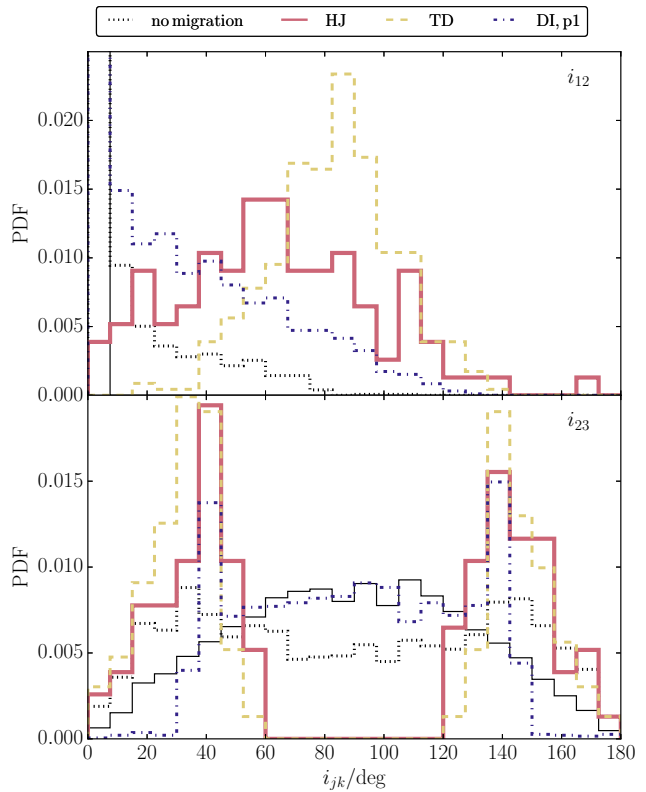}
\caption{\small The distributions of the inclinations $i_{12}$ (top panel) and $i_{23}$ (bottom panel), arranged by simulation outcomes and at the end of the simulations. }
\label{fig:mutual_inclination_distributions_test03.eps}
\end{figure}

In addition to the semimajor axes, there are distinctive features of the HJ systems with respect to the inclinations. In \F\,\ref{fig:mutual_inclination_distributions_test03.eps}, we show the distributions of the inclinations $i_{12}$ and $i_{23}$ in the top and bottom panels, respectively, arranged by simulation outcomes and at the end of the simulations (or until a tidal disruption or dynamical instability occurred). The $i_{12}$ distribution is broadly distributed between $0^\circ$ and $\approx 140^\circ$. Qualitatively, one might expect peaks near $\sim 40^\circ$ and $130^\circ$ in the classical picture of high-$e$ LK migration. In this case, however, the second planet continues to change its orientation after the innermost planet became a HJ, which tends to isotropize the $i_{12}$ distribution. 

The distribution of $i_{23}$ for HJ systems, on the other hand, is strongly peaked near $40^\circ$ and $140^\circ$. This can be understood as follows. If $i_{23}$ is high, i.e. in the LK regime between $40^\circ$ and $130^\circ$, then orbit 2 can become highly eccentric due to the secular torque of orbit 3, likely triggering a dynamical instability. This is reflected in the bottom panel of \F\,\ref{fig:mutual_inclination_distributions_test03.eps}: the dynamical instability systems show a strong preference for this inclination window. Only if $i_{23}$ is outside of this window, the planets can remain dynamically stable whereas $i_{23}$ is still large enough to drive a high mutual inclination between their orbits.

Based on these preferences, we ran additional simulations aimed at increasing the HJ fractions by restricting the parameter ranges as follows: $1\,\mathrm{AU}<a_1<4\,\mathrm{AU}$, $30\,\mathrm{AU}<a_2<50\,\mathrm{AU}$, $200 \,\mathrm{AU} \lesssim a_3 \lesssim 1000 \, \mathrm{AU}$, $30^\circ< i_{23} < 50^\circ$ and $130^\circ < i_{23} < 150^\circ$. This restricted setup corresponds to a fraction of $\approx 0.045$ of the initial population synthesis setup. The HJ fraction for the restricted choices is $\approx 0.23$ for $t_\mathrm{V,1}=0.014\,\mathrm{yr}$, about ten (four) times larger compared to the unrestricted runs with (without) the second planet. The tidal disruption fraction in the restricted run is $\approx 0.31$, and $\approx 0.12$ of systems became dynamically unstable. Further fine-tuning would likely increase the HJ fraction even more. For comparison, in the case of no second planet and these restricted parameters, we found no HJs and only non-migrating systems.

\section{Discussion}
\label{sect:discussion}
\subsection{Expected properties of the hidden companion}
We have shown that the addition of a second planet in a stellar binary system can strongly enhance the rate of HJs formed through high-$e$ migration in binaries that are otherwise not conducive to HJ migration. There are specific constraints on the properties of this second planet; most importantly, that the ratio $\mathcal{R}$ in equation~(\ref{eq:R}) be close to unity (within a factor of a few, based on \F\,\ref{fig:arranged_sma_distributions_combined_test03.eps}). Neglecting planetary-to-stellar masses, this gives the estimate
\begin{align}
\label{eq:a_2_constr}
\nonumber a_2 &\sim \mathcal{R}^{1/3} (a_1 a_3^2)^{1/3} \left ( \frac{m_2}{m_3} \right)^{2/9} \left ( \frac{1-e_3^2}{1-e_2^2} \right )^{1/3} \\
\nonumber &\approx 43 \, \mathrm{AU} \, \left ( \frac{a_1}{2\,\mathrm{AU}} \right )^{1/3} \left ( \frac{a_3}{500\,\mathrm{AU}} \right )^{2/3} \left ( \frac{m_2}{5 \, M_\mathrm{J}} \right )^{2/9} \\
&\quad \times \left ( \frac{m_3}{0.4 \, \mathrm{M}_\odot} \right )^{-2/9} \left ( \frac{1-e_3^2}{1-e_2^2} \right )^{1/3},
\end{align}
where we set $\mathcal{R}=3$ for the numerical value. This equation can be used to estimate the semimajor axis of a hypothetical companion planet with mass $m_2$ to a HJ in a stellar binary with semimajor axis $a_3$ and eccentricity $e_3$, assuming an initial proto-HJ semimajor axis $a_1$. The dependence on the eccentricities is weak, unless the eccentricities are implausibly high ($\gtrsim 0.9$). In addition to this condition to be approximately fulfilled, we also expect that currently, the HJ is highly inclined with respect to the second planet (cf. the top panel of \F\,\ref{fig:mutual_inclination_distributions_test03.eps}), whereas the second planet has a lower inclination with respect to the stellar binary, peaked around $40^\circ$ or $140^\circ$ (cf. the bottom panel of \F\,\ref{fig:mutual_inclination_distributions_test03.eps}).

\subsection{Dynamical stability outcomes}
As mentioned in \S\,\ref{sect:pop_syn:results}, the two planets become unstable in a large fraction ($\sim 0.5$) of the simulations (using the criterion of \citealt{2015ApJ...808..120P}). These instabilities are triggered by an increase of the eccentricity of the orbit of the second planet by the secular torque of the stellar binary companion, and most likely result in ejections of planets (particularly if $m_1 > m_2$) or collisions of planets with the primary star (particularly if $m_1 < m_2$). In principle, planets could also be tidally captured by one of the stars, producing HJs. This model is beyond the scope of this paper, but merits further investigation.

\subsection{Planet-planet perturbations in absence of the stellar binary companion}
Even in the absence of the stellar binary companion, the second planet could (secularly) perturb the proto-HJ and drive high-$e$ migration. This model was considered by \citet{2015ApJ...805...75P}. To investigate this possibility in the context of our simulations, we also carried out a population synthesis run with the stellar binary companion removed, but with the other properties unchanged, i.e. with the planets initially in coplanar and (nearly) circular orbits. The result is that the eccentricities change only very little, and no HJs are produced. This can be understood from the circular and coplanar orbits, and the relatively large orbital separations between the planets in our simulations (cf. the top panel of \F\,\ref{fig:arranged_sma_distributions_combined_test03.eps}).

\section{Conclusions}
\label{sect:conclusions}
Observed correlations of HJ and stellar binary companion fractions suggest a link between HJ formation and the presence of a binary companion. One possibility is that the stellar binary companion induces Lidov-Kozai (LK) oscillations in the proto-HJ orbit,  thought to have an initial semimajor axis of a few AU, triggering orbital dissipation due to tides. However, a major problem of this model is that the companions are typically too wide to induce high-eccentricity LK oscillations due to suppression by short-range forces. 

We have proposed a modification in which there is a second planet orbiting in-between the proto-HJ and the stellar binary companion, at a few tens of AU from the primary star. If the planets were formed in the same disk, then their orbits were initially likely (close to) coplanar. Over time, a large mutual inclination could be induced due to the secular torque of the stellar binary companion. Consequently, the second planet could drive LK oscillations in the proto-HJ orbit, causing the latter to migrate. 

Our model predicts a high occurrence rate of planetary companions to HJs in stellar binaries. In order for the process to be effective, the semimajor axis of the second planet should lie in a specific range given the planetary mass and stellar binary semimajor axis. The required typical planetary companion semimajor axis can be estimated from the simple analytical relation in equation~(\ref{eq:a_2_constr}). In addition, we require the inclination of the second planet orbit with respect to the stellar binary to be near $40^\circ$ or $140^\circ$ to avoid triggering dynamical instabilities by high eccentricities of the orbit of the second planet, yet still inducing a large mutual inclination between the two planets. By applying these restrictions, the HJ fractions in our population synthesis simulations were enhanced by a factor of $\sim 10$. However, the restricted systems correspond to only a small fraction, $\approx 0.045$, of the systems sampled without additional restrictions, indicating that the parameter space for HJ enhancement is small.

A giant planet companion to a HJ with a semimajor axis of $\sim 40\,\mathrm{AU}$ in a stellar binary should be detectable by future radial velocity surveys, which are currently limited to $\sim$ 20 AU for masses of $\sim1 \,M_\mathrm{J}$ \citep{2016ApJ...821...89B}. Also, AO contrast curves can place upper limits on the semimajor axis and mass. Furthermore, Gaia astrometry might constrain companions of several $M_\mathrm{J}$ at $\sim 40\,\mathrm{AU}$ by resolving a section of the planetary orbit within Gaia's mission time \citep{2014ApJ...797...14P}. 

Should these companions turn out to be absent, then this would strongly suggest that the correlation between the HJ and stellar binary companion occurrence rates is due to another phenomena, e.g. planet formation processes unrelated to secular dynamical effects. It might be the final nail in the coffin for high-$e$ migration to be the dominant contribution to the formation of HJs in stellar binaries.

\section*{Acknowledgements}
We thank Dong Lai, Heather Knutson and Henry Ngo for stimulating discussions and comments on the manuscript, and the anonymous referee for an insightful report. ASH gratefully acknowledges support from the Institute for Advanced Study.

\bibliographystyle{apj}
\bibliography{literature}

\end{document}